\documentclass{ws-ijgmmp}
\usepackage[colorlinks=true]{hyperref}
\usepackage{graphicx}

\begin{document}

\markboth{Dagmar Läänemets, Manuel Hohmann, Christian Pfeifer}{Observables from spherically symmetric modified dispersion relations}

%
\catchline{}{}{}{}{}
%

\title{Observables from spherically symmetric modified dispersion relations}

\author{Dagmar Läänemets}
\address{Laboratory of Theoretical Physics, Institute of Physics, University of Tartu, W. Ostwaldi 1, 50411 Tartu, Estonia.\\
\email{dagmar.laanemets@gmail.com}}

\author{Manuel Hohmann}
\address{Laboratory of Theoretical Physics, Institute of Physics, University of Tartu, W. Ostwaldi 1, 50411 Tartu, Estonia.\\
\email{manuel.hohmann@ut.ee}}

\author{Christian Pfeifer}
\address{ZARM, University of Bremen, Am Fallturm 2, 28359 Bremen, Germany.\\
\email{christian.pfeifer@zarm.uni-bremen.de}}

\maketitle

\begin{history}
\received{(Day Month Year)}
\revised{(Day Month Year)}
\end{history}

\begin{abstract}
In this work we continue the systematic study of observable effects emerging from modified dispersion relations. We study the motion of test particles subject to a general first order modification of the general relativistic dispersion relation as well as subject to the $\kappa$-Poincar\'e dispersion relation in spherical symmetry. We derive the corrections to the photon sphere, the black hole shadow, the Shapiro delay and the light deflection and identify the additional dependence of these observables on the photons' four momentum, which leads to measurable effects that can be compared to experimental data. The results presented here can be interpreted in two ways, depending on the origin of the modified dispersion relation: on the one hand as prediction for traces of quantum gravity, when the modified dispersion relation is induced by phenomenological approaches to quantum gravity, on the other hand as predictions of observables due to the presence of a medium, like a plasma, which modifies the dispersion relation of light on curved spacetimes.
\end{abstract}


\section{Introduction}
Dispersion relations are constraints which the relativistic momentum of a particle has to satisfy in order to be physically viable and they determine the particle's trajectory. In the case of an underlying physical field theory, they can be derived from the field equations of a theory in terms of the point particle (or geometric optical) limit. Moreover, they emerge as Casimir operator of symmetries one imposes on relativistic point particle kinematics, or are reconstructed by comparing a phenomenological ansatz with measurements.

In the standard model of particle physics on curved spacetime and general relativity, which are based on local Lorentz invariance, the point particle limit of the field equations lead to the general relativistic dispersion relation of free massive and massless point particles. The same result is obtained by imposing (local) Lorentz invariance as symmetries of relativistic point particle kinematics.

Modifications of the general relativistic dispersion relation, or short modified dispersion relations (MDRs), have wide applications in physics. Besides arising from modified gravity theories~\cite{CANTATA:2021ktz}, they are used in quantum gravity phenomenology~\cite{AmelinoCamelia:2008qg,Addazi:2021xuf} to describe the influence of the quantum nature of gravity on fields and particles. Most prominent models, which introduce an additional observer independent scale to the speed of light, are doubly or deformed special relativity (DSR)~\cite{AmelinoCamelia:2000mn, AmelinoCamelia:2000ge,KowalskiGlikman:2002we}, more generally deformed relativistic kinematics~\cite{Carmona:2019oph, Carmona:2019fwf, Carmona:2012un, Pfeifer:2021tas}, and non-commutative spacetime geometries~\cite{Lukierski:1991pn,Majid:1994cy,Snyder:1946qz,Lobo:2016blj,Deriglazov:2019vcj}. Furthermore, MDRs describe the propagation of particles and fields on quantum spacetimes constructed in string theory~\cite{Ellis:1999sd,AmelinoCamelia:1996pj} or Loop quantum gravity~\cite{Assaniousssi:2014ota,Amelino-Camelia:2016gfx,Brahma:2018rrg,Lobo:2019jdz}. Moreover, MDRs describe the point particle limit of the Standard-Model Extension (SME)~\cite{Bluhm:2005uj}, which studies the consequences of adding non local Lorentz invariant terms to the standard model of particle physics, and the point particle limit of effective field theories which describe the propagation of fields through media~\cite{Klimes2002,Cerveny2002,Yajima1763,Antonelli2003,Gibbons:2011ib,Rubilar:2007qm,Punzi:2007di,Perlick}. A nice and comprehensive overview on the relation between field theories and dispersion relation is given in the article~\cite{Raetzel:2010je}.

Modified dispersion relations lead to a momentum dependent spacetime geometry, which can be derived in terms of what mathematicians call Hamilton geometry~\cite{Miron,Barcaroli:2015xda,Pfeifer:2021tas}, or its dual formulation, in terms of a velocity dependent Finsler spacetime geometry~\cite{Girelli:2006fw,Amelino-Camelia:2014rga,Pfeifer:2019wus}. Depending on the question to be investigated, the momentum dependent or the velocity dependent representation of the spacetime geometry, or switching between both, turns out to be more convenient to study the properties and observable predictions of a MDR.

Observable consequences of MDRs are often studied in highly symmetric situations.

Many investigations in the literature focus on flat spacetimes, i.e.\ modifications of the special relativistic dispersion relation. In this case modifications of the classical special relativistic effects like length contraction and time dilation are investigated~\cite{Gurlebeck:2018nme,Lobo:2020qoa} as well as particle physics processes from deformed relativistic kinematics or quantum field theories~\cite{Carmona:2000ze,Kostelecky:2008ts,Kostelecky:2009zp,Fewster:2017mtt,Carmona:2019xxp,Arzano:2019toz,Lobo:2021yem}.

One of the most studied features of MDRs is derived for spatially homogeneous and isotropic cosmological symmetry: a possible four-momentum dependent time delay in the time of arrival of high energetic photons from gamma-rays that are emitted at the same time at large distance. For MDRs that are polynomial in the energy of the photons such time delays have been presented in the articles~\cite{Ellis:2002in,Ellis:2005wr,Jacob:2008bw}. The derivation has been extended to numerous DSR models~\cite{AmelinoCamelia:2012it,Rosati:2015pga} and general first order MDRs~\cite{Pfeifer:2018pty}, including the $\kappa$-Poincar\'e dispersion relation in the bicrossproduct basis~\cite{Barcaroli:2016yrl}. Currently, searches for such time delays in the data of the H.E.S.S., MAGIC and VERITAS telescopes are in progress~\cite{Terzic:2021rlx,Bolmont:2022yad}.

Besides flat spacetimes and spatially homogeneous and isotropic cosmological symmetry only a few observables for MDRs have been derived. First results on photon orbits and lensing for the $\kappa$-Poincar\'e dispersion relation in the bicrossproduct basis have been made~\cite{Barcaroli:2017gvg,Glicenstein:2019rzj}.

In this article we derive systematically observables from the $\kappa$-Poincar\'e dispersion relation in the bicrossproduct basis and general first order MDRs in spherical symmetry. In particular we will derive: the photon orbits, which are the key ingredient for the prediction of the shadow of black holes~\cite{Perlick:2021aok}, an observable nowadays accessible thanks to the Event Horizon telescope~\cite{EventHorizonTelescope:2019dse}; the Shapiro delay; and the deflection angle of light, which is a first step towards a general gravitational lensing formula for MDRs. We find that generically these classical general relativistic observables become dependent on the energy or angular momentum parameter of the photon trajectory. For a given MDR model, our results determine the precise quantitative functional dependence on these parameters. Hence the observation of the black hole shadow, the Shapiro delay or light deflection (lensing) in different frequencies, and the comparison of these observations can distinguish and constrain the different MDRs.

This work continues the exploration of observables from MDRs and the search for constraints or evidence for traces of quantum gravity in a not yet explored regime. In addition, the results for the first order MDRs are also valid for spherical symmetric configurations of media, like plasmas, on curved spacetime.

The presentation of the results in this article is organized as follows. We begin by discussing how dispersion relations can be interpreted as Hamilton functions on the point particle phase space of a curved spacetime manifold in Section~\ref{sec:Ham}. In Section~\ref{sec:sphdisp} we recall the form of the most general spherically symmetric dispersion relation and use these insights to display the most general spherically symmetric  $\kappa$-Poincar\'e dispersion relation and first order MDRs. The main results of this article are presented in section~\ref{sec:obs} where we present the explicit expressions for the photon orbits and black hole shadows (Section~\ref{ssec:photonorb}), the Shapiro delay  (Section~\ref{ssec:shapiro}) and the light deflection angle  (Section~\ref{ssec:light}). We conclude this article in Section~\ref{sec:conc}.

\section{Dispersion relations as Hamilton functions}\label{sec:Ham}
Point particle dispersion relations on a curved spacetime manifold $M$ can be understood as being defined by Hamilton functions $H(x,p)$ on the cotangent bundle $T^*M$ (physically speaking position and momentum space) of spacetime~\cite{Barcaroli:2015xda}.

Throughout this article we consider a four dimensional spacetime $M$ and the cotangent bundle in manifold induced coordinates. This means an element $P\in T^*_xM\subset T^*M$ can be expanded in local coordinates of the base manifold $M$ as $P= p_\mu \mathrm{d} x^\mu$ and will be identified with the tuple $(x,p)$. Differential operators will be abbreviated as $\partial_\mu = \frac{\partial}{\partial x^\mu}$ and $\bar\partial^\mu = \frac{\partial}{\partial p_\mu}$.

In these local coordinates a Hamilton function $H:T^*M \rightarrow \mathbb{R}$ defines a dispersion relation via
\begin{align}
	H(x,p) = -m^2.
\end{align}
Moreover it determines the trajectories for point particles via the usual Hamilton equations of motion
\begin{align}
	\dot p_\mu = - \partial_\mu H,\quad \dot x^\mu = \bar\partial^\mu H\,,
\end{align}
where a dot above the symbol denotes a derivative with respect to an arbitrary curve parameter $\tau$. Geometrically they can be related to the autoparallels of an in general non-linear connection on the cotangent bundle which is determined from the Hamilton function. For the analysis of point particle motion in spherical symmetry subject to a MDR the whole geometric framework of Hamilton or generalized Hamilton geometry is not needed. For the interested reader we refer for the full geometric framework to the articles~\cite{Pfeifer:2021tas,Miron,Barcaroli:2015xda}.

The Hamilton functions whose phenomenology we will study in this article are on the one hand general perturbations of the Hamilton function of freely falling particles in general relativity
\begin{align}
	H_\epsilon(x,p) = \frac{1}{2}g_x(p,p) + \epsilon h(x,p)\,,
\end{align}
where $g_x(p,p)  = g^{\mu\nu}(x)p_\mu p_\nu$ is the metric norm of a particle's momentum $p$, $h(x,p)$ is a perturbation function and $\epsilon$ is a perturbation parameter, that can be identified with the quantum gravity scale, the Planck scale $\ell$, in the context of quantum gravity phenomenology~\cite{AmelinoCamelia:2008qg,Addazi:2021xuf}, or with a parameter that characterizes the interaction of particles with a medium when one studies the propagation of light through media like a plasma or a crystal. The units of $\epsilon h(x,p)$ must match the units of $g_x(p,p)$, which units these are depends on the application under consideration. For a medium with refractive index $n(x,p)$  the general Hamiltonian, which describes light propagation in the moving media on curved spacetime is~\cite{Synge}
\begin{align}
	H(x,p) = \frac{1}{2}\left(g_x(p,p) - (n(x,p)^2-1)V_x(p)^2\right)\,,
\end{align}
where $V_x(p) = V^\mu(x)p_\mu$ is the $4$-velocity of the medium at the position $x$ contracted with the particle's momentum. For media with a weak interaction between light and the medium one may expand the refractive index as $n(x,p)^2 = 1 + \epsilon f(x,p)$ which yields a perturbation function of the form
\begin{align}\label{eq:pertmedium}
	h(x,p) = - \frac{1}{2}f(x,p)V_x(p)^2\,.
\end{align}

On the other hand we consider the $\kappa$-Poincar\'e dispersion relation in bicrossproduct basis on curved spacetime~\cite{Barcaroli:2017gvg} to all orders
\begin{align}\label{eq:kappadisp}
	H_\ell(x,p)
	&= -\frac{2}{\ell^2}\sinh\left(\frac{\ell}{2}Z_x(p)\right)^2 +\frac{1}{2} e^{\ell Z_x(p)}(g_x(p,p) + Z_x(p)^2) \,.
\end{align}
It is defined in terms of the metric norm of the particle's momenta $g_x(p,p)$ and a unit timelike vector field acting on the momentum $Z_x(p) = Z^\mu(x) p_\mu$, where $g_{\mu\nu}  Z^\mu Z^\nu =-1$. It is motivated in the context of quantum gravity phenomenology and deformed relativistic kinematics obtained from quantum deformations of the Poincar\'e algebra~\cite{Lukierski:1991pn,AmelinoCamelia:2008qg,Barcaroli:2017gvg, Addazi:2021xuf}.

For the analysis of particle motion in spherical symmetry in the next sections it is convenient to use spherical coordinates $(t,r,\theta,\phi, p_t,p_r,p_\theta,p_\phi)$ for which the Hamilton equations of motion become
\begin{align}
	\dot t &=\bar\partial^t H && \dot p_t = - \partial_t H \label{eq:t}\\
	\dot r &=\bar\partial^r H && \dot p_r = -\partial_r H\\
	\dot \theta &=\bar\partial^\theta H && \dot p_\theta = -\partial_\theta H \label{eq:theta}\\
	\dot \phi &=\bar\partial^\phi H && \dot p_\phi = -\partial_\phi H \label{eq:phi}\,.
\end{align}

To derive the innermost circular photon orbits in section \ref{ssec:photonorb} we need to solve the $\dot r = 0$ and $\ddot r = 0$, while the Shapiro delay will be derived in section \ref{ssec:shapiro} from the quotient $\dot t/\dot r$ and the light deflection from the quotient $\dot \phi/\dot r$ in section \ref{ssec:light}.

Before we derive these observables we derive the Hamilton equations of motion for general spherically symmetric Hamilton functions (dispersion relations).

\section{Static spherically symmetric dispersion relations}\label{sec:sphdisp}
Spherically symmetry of a dispersion relation is implemented by demanding that the Hamilton function $H(x,p)$ is invariant along the flow of the symmetry generating vector fields which form the $\mathfrak{so}(3)$ algebra. In coordinates $(t,r,\theta,\phi, p_t,p_r,p_\theta,p_\phi)$  they take the form
\begin{align}
	X_1 = \sin\varphi\partial_{\vartheta} + \frac{\cos\varphi}{\tan\vartheta}\partial_{\varphi}\,,
	\quad
	X_2 = -\cos\varphi\partial_{\vartheta} + \frac{\sin\varphi}{\tan\vartheta}\partial_{\varphi}\,,
	\quad
	X_3 = \partial_\varphi\,.
\end{align}
Solving the Killing equation for Hamilton functions, as it has been done in~\cite{Barcaroli:2017gvg}, yields that a spherically symmetric Hamilton function cannot be an arbitrary function of the coordinates  $(t,r,\theta,\phi, p_t,p_r,p_\theta,p_\phi)$, but must be of the form
\begin{equation}\label{eq:Hsph}
	H(t,r,\theta,\phi, p_t,p_r,p_\theta,p_\phi) = H(t,r,p_t,p_r,w),\quad w^2 = w(\theta,p_\theta,p_\phi)^2 = p_\theta^2 + \frac{1}{\sin\theta^{2}} p_\phi^2\,.
\end{equation}
In the following we are interested in stationary configurations, so we drop the $t$-dependence of the Hamiltonian.

For such static spherically symmetric Hamilton functions $H(r,p_t,p_r,w)$ the usual general conclusions on particle motion can be drawn. There exist two constants of motion along the solution of the Hamilton equations of motion, the energy $p_t = \mathcal{E}$ which can be obtained from the $p_t$-equation in \eqref{eq:t}, and the angular momentum $p_\phi = \mathcal{L}$ which follows immediately from the $p_\phi$-equation in \eqref{eq:phi}. Moreover, without loss of generality, the motion of particles can be studied in the equatorial plane, since the $\theta$-equations \eqref{eq:theta}
\begin{align}
	\dot \theta &=\bar\partial^\theta H = \frac{\partial H}{\partial w}\frac{1}{2w} \bar\partial^\theta w^2 = \frac{\partial H}{\partial w}\frac{1}{w}  p_\theta  \\
	\dot p_\theta &= -\partial_\theta H = - \frac{\partial H}{\partial w}\frac{1}{2w} \partial_\theta w^2 = \frac{\partial H}{\partial w}\frac{1}{w} \frac{\cos \theta}{ \sin^3 \theta} p_\phi^2\,,
\end{align}
are solved by setting $\theta = \frac{\pi}{2}$ and $p_\theta = 0$. Thus, along the trajectories or particles propagating along the solutions of the Hamilton equations of motion we have that $w(\pi/2,0,p_\phi)^2 = p_\phi^2 = \mathcal{L}^2$, meaning that orbital motion is planar.

The remaining equations from \eqref{eq:t} to \eqref{eq:phi}, from which the desired observables will be derived in the following sections are the $t$, $r$, $p_r$ and $\phi$ equations. They become
\begin{align}
	\dot t &=\bar\partial^t H(r,\mathcal{E},p_r,\mathcal{L}) \,,\label{eq:dottsph}\\
	\dot \phi &= \bar\partial^\phi H(r,\mathcal{E},p_r,\mathcal{L}) = \partial_w H(r,\mathcal{E},p_r,\mathcal{L})\,,\label{eq:dotphisph}\\
	\dot r &= \bar\partial^r H(r,\mathcal{E},p_r,\mathcal{L})\,,\label{eq:dotrsph}\\
	\dot p_r &= - \partial_r H(r,\mathcal{E},p_r,\mathcal{L})\label{eq:dotprsph}\,.
\end{align}

Before we derive the observables we discuss the spherically symmetric $\kappa$-Poincar\'e dispersion relation and spherically symmetric first oder MDRs.
\subsection{General first order modified dispersion relations}\label{ssec:sph1stOrder}
General static first order modifications of the static general relativistic point particle Hamiltonian in spherical symmetry are of the form
\begin{equation}\label{eq:mdrpert}
	H_\epsilon(x,p) = \frac{1}{2}\left( -a p_t^2 + b p_r^2 + \frac{1}{r^2} w^2 \right) + \epsilon h(r,p_t,p_r,w)\,.
\end{equation}
where $a=a(r)$ and $b=b(r)$.

The Hamilton equations of motion become
\begin{align}
	\dot t  &= - a \mathcal{E} + \epsilon \bar\partial^t h(r,\mathcal{E}, p_r,\mathcal{L})\,,\label{eq:dottpert}\\
	\dot \phi &= \frac{\mathcal{L}}{r^2} + \epsilon \partial_w h(r,\mathcal{E}, p_r,\mathcal{L})\,,\label{eq:dotphipert}\\
	\dot r &= b p_r + \epsilon \bar\partial^r h(r,\mathcal{E}, p_r,\mathcal{L})\,,\label{eq:dotrpert}\\
	\dot p_r &= \frac{1}{2} a' \mathcal{E}^2 - \frac{1}{2} b' p_r^2 + \frac{\mathcal{L}^2}{r^3} -\epsilon \partial_r h(r,\mathcal{E}, p_r,\mathcal{L})\label{eq:dotprpert}\,,
\end{align}
where a prime $'$ denotes derivative with respect to $r$.

\subsection{The $\kappa$-Poincar\'e case}\label{ssec:sphkappa}
In spherical symmetry the static $\kappa$-Poincar\'e Hamiltonian \eqref{eq:kappadisp} becomes, see~\cite{Barcaroli:2017gvg},
\begin{align}\label{eq:genkappasph}
	H_\ell = - \frac{2}{\ell^2}\sinh \left(\frac{\ell}{2} (c p_t + d p_r) \right)^2 + \frac{1}{2} e^{\ell (c p_t + d p_r)} \left( (-a + c^2)p_t^2 + 2 c d p_r p_t + (b+d^2)p_r^2 + \frac{w^2}{r^2}\right)\,,
\end{align}
where $c=c(r)$ and $d=d(r)$ are the components of the vector field $Z = c\partial_t + d \partial_r$ and $a=a(r)$ and $b=b(r)$ are the component of the spherically symmetric metric which appear in $g_x(p,p) = -a p_t^2 + b p_r^2 + w^2/r^2$. The components are related by the constraint that $Z$ should be unit timelike with respect to $g$
\begin{align}
	-\frac{c^2}{a} + \frac{d^2}{b} = -1\, \Rightarrow c^2 = a \left(\frac{d^2}{b} + 1\right)\,.
\end{align}
To study phenomenological effects of the $\kappa$-deformation of Schwarzschild geometry explicitly we set $b=a^{-1} = (1-\frac{r_s}{r})$, $d=0$ and thus $c=\sqrt{a}$, which yields
\begin{align}\label{eq:kappaschw}
	H_\ell = - \frac{2}{\ell^2}\sinh \left(\frac{\ell}{2} \sqrt{a} p_t \right)^2 +\frac{1}{2} e^{\ell \sqrt{a} p_t}\left(  \frac{p_r^2}{a} + \frac{w^2}{r^2}\right)\,.
\end{align}
For this Hamiltonian the point particle equations of motion \eqref{eq:dottsph} to \eqref{eq:dotprsph} become
\begin{align}
	\dot t &= - \frac{1}{\ell} \sqrt{a} \sinh\left( \ell\mathcal{E} \sqrt{a}\right) +\frac{1}{2} \ell \sqrt{a} e^{\ell \sqrt{a}\mathcal{E}} \left(\frac{p_r^2}{a} + \frac{\mathcal{L}^2}{r^2}\right)\,,\label{eq:dottsphkappa}\\
	\dot \phi &=  e^{\ell\sqrt{a} \mathcal{E}} \frac{\mathcal{L}}{r^2}\,,\label{eq:dotphisphkappa}\\
	\dot r &=  e^{\ell\sqrt{a} \mathcal{E}} \frac{p_r}{a}\,,\label{eq:dotrsphkappa}\\
	\dot p_r &= \frac{ \mathcal{E} a'}{\ell2  \sqrt{a}} \sinh\left( \ell \mathcal{E} \sqrt{a}\right)
	+ \frac{1}{2}e^{\ell\sqrt{a} \mathcal{E}}
	\left( \frac{a'}{a^2} p_r^2 + \frac{2\mathcal{L}^2}{r^3} - \frac{\ell a'}{2 \sqrt{a}} \mathcal{E} \left(\frac{p_r^2}{a} + \frac{\mathcal{L}^2}{r^2}\right)\right) \label{eq:dotprsphkappa}\,,
\end{align}
again a prime $'$ denotes a derivative with respect to $r$.

In the language of the first order perturbations discussed in the previous section (setting $\epsilon=\ell$) the spherically symmetric $\kappa$-Poincar\'e dispersion relation \eqref{eq:genkappasph} leads to a perturbation function
\begin{align}
	h(r,p_t,p_r,w) =  \frac{1}{2}(c p_t + d p_r) \left(- a p_t^2 + b p_r^2 + \frac{1}{r^2}w^2 + (c p_t + d p_r)^2 \right)\,,
\end{align}
For the models \eqref{eq:kappaschw} the perturbation reduces to
\begin{align}\label{eq:kappaschwpert}
	h(r,p_t,p_r,w) =  \frac{1}{2} p_t \sqrt{a} \left(\frac{p_r^2}{a} + \frac{1}{r^2}w^2 \right)\,.
\end{align}

\section{Observables}\label{sec:obs}
In this section we will derive the innermost circular photon orbits, i.e. the photon shell which is the boundary of the shadow of a black hole, the Shapiro time delay and the light deflection angle for the $\kappa$-Poincar\'e dispersion relation as well as for general first order modifications of the general relativistic dispersion relation.

\subsection{Circular photon orbits}\label{ssec:photonorb}
In spherically symmetric spacetime geometries exist circular photon orbits, which form the so called photon sphere. The radius of the photon sphere is the closest radial distance a photon can approach to a black hole before it falls inevitably in and will disappear behind the event horizon. The angular size of the shadow of black holes is determined by this critical radius and the radial position of the observer~\cite{Perlick:2021aok, Grenzebach:2014fha,Fuller:2019sxi}, as we will recall at the end of this subsection. For generic MDRs the photon sphere will not be universal for all photons, but depends on their energy or angular momentum. Hence any evidence for a frequency dependence of the shadow of a black hole would be an observational signal for a MDR.

Circular orbits are characterized by the fact that $\dot r=0$, and hence \eqref{eq:dotrsph} implies that $\bar\partial^rH=0$. This equation can be used to determine $p_r = p_r(r, \mathcal{E},\mathcal{L})$. The next step is to use this result in the dispersion relation $H(r,\mathcal{E},p_r(r, \mathcal{E},\mathcal{L}),\mathcal{L}) = 0$ for photons and to solve for $\mathcal{E}(r,\mathcal{L})$. Finally the equation $\dot p_r = -\partial_rH$ can be used together with the already obtained result to find the innermost circular orbit of photons as $r=r(\mathcal{L})$.

\subsubsection{General first order modified dispersion relations}
To calculate the influence of a general first order MDR \eqref{eq:mdrpert} on innermost circular photon orbits we consider photon trajectories $c(\tau) =  (x(\tau),p(\tau)) $ in phase space of the type
\begin{align}
	c(\tau) = (x^\mu(\tau),p_\mu(\tau)) = (x^\mu_0(\tau),p_{\mu0}(\tau)) + \epsilon (x^\mu_1(\tau),p_{\mu1}(\tau))\,.
\end{align}
The zeroth order is a solution of the Hamilton equations of motion for a relativistic particle and the first order is the modification induced by the modification $h$ of the general relativistic Hamilton function. From our general considerations this also means that the energy and the angular momentum can be split into zeroth and first order terms
\begin{align}\label{eq:ELpert}
	\mathcal{E} = \mathcal{E}_0 +\epsilon \mathcal{E}_1,\quad \mathcal{L} = \mathcal{L}_0 +\epsilon \mathcal{L}_1\,.
\end{align}

For the circular orbits, again, we first consider $\dot r=0$, which implies from \eqref{eq:dotrpert}
\begin{align}\label{eq:prpert3}
	p_r = -\frac{\epsilon}{b} \bar\partial^r h(r,\mathcal{E}, p_r,\mathcal{L})\,.
\end{align}
Since also $\dot p_t=\dot p_\phi=0$, we can take another $\tau$ derivative to find
\begin{align}\label{eq:prpert4}
	\dot p_r = -\frac{\epsilon}{b} \bar{\partial}^r\bar\partial^r h(r,\mathcal{E},p_r,\mathcal{L})\dot p_r\,.
\end{align}
Evaluating the equations \eqref{eq:prpert3} and \eqref{eq:prpert4} order by order implies
\begin{align}\label{eq:prpertall}
	p_{r0} = 0,\quad p_{r1} = \frac{\epsilon}{b(r_0)}\bar\partial^r h(r_0,\mathcal{E}_0,0,\mathcal{L}_0) = \frac{\epsilon}{b_0}\bar\partial^r h_0,\quad  \dot p_{r1}=0\,.
\end{align}
Functions labeled with an index $0$ are evaluated at the general relativistic value of their argument.

Next, we can use the dispersion relation \eqref{eq:mdrpert} to express either $\mathcal{E}$ or $\mathcal{L}$ in terms of each other.We decide to eliminate $\mathcal{E}$ by
\begin{align}\label{eq:E1}
	\mathcal{E}^2 = \frac{1}{a} \left( \frac{\mathcal{L}^2}{r^2} + 2m^2 + p_r^2 b \right) + \epsilon \frac{2}{a}h(r,\mathcal{E}, p_r,\mathcal{L})\,.
\end{align}
Thus, using \eqref{eq:E1} and \eqref{eq:prpertall} in the remaining equation \eqref{eq:dotprpert} gives
\begin{align}\label{eq:prpert2}
	0 = \frac{1}{2} \frac{a'}{a}\left( \frac{\mathcal{L}^2}{r^2} + 2m \right)  + \frac{\mathcal{L}^2}{r^3} + \epsilon \left(\frac{a'}{a}h(r,\mathcal{E},p_r,\mathcal{L}) - \partial_r h(r,\mathcal{E}, p_r,\mathcal{L})\right)\,.
\end{align}
To solve this equation for the correction of the radius of the circular particle orbits order by order, we now decompose all quantities into their background value and the correction due to the MDR $r = r_0 + \epsilon r_1$, likewise $\mathcal{L} = \mathcal{L}_0 + \epsilon \mathcal{L}_1$ and $m = m_0 + \epsilon m_1$. To first order in $\epsilon$ we find
\begin{align}
	0 &= \frac{1}{2} \frac{a'_0}{a_0}\left( \frac{\mathcal{L}^2_0}{r^2_0} + 2m_0^2 \right) + \frac{\mathcal{L}^2_0}{r_0^3}\\
	&+ \epsilon \bigg[  \left(  \frac{\mathcal{L}_0^2}{2} \left(\frac{ a''_0}{r^2_0 a_0} - \frac{a'^2}{r^2_0 a_0^2}- \frac{2}{r^3_0}\frac{a'_0}{a_0} -  \frac{6}{r^4_0} \right) - m_0^2 \left(\frac{a'^2}{a_0^2} - \frac{a''_0}{a_0} \right) \right) r_1\\
	&+ \left(\frac{a'_0}{a_0}h_0 - \partial_r h_0\right) + 2 \frac{a'_0}{a_0} m_0 m_1 + \left (\frac{a'_0}{a_0 r_0^2} + \frac{2}{r_0^3} \right) \mathcal{L}_0 \mathcal{L}_1\bigg]\,.
\end{align}
The zeroth order, i.e.\ the general relativistic value of the circular orbits is determined by the equation
\begin{align}\label{eq:r0}
	\frac{1}{2} \frac{a'_0}{a_0}\left( \frac{\mathcal{L}_0^2}{r^2_0} + 2m_0^2 \right) + \frac{\mathcal{L}_0^2}{r_0^3} = 0\,,
\end{align}
while, for the first order correction to the circular orbits of massive and massless particles we obtain
\begin{align}\label{eq:r1}
	r_1 &= \frac{  r_0^3 (\partial_r h_0 - \frac{a'_0}{a_0}(h_0  + 2 m_0 m_1)) - \left (\frac{a'_0}{a_0 } r_0 + 2 \right) \mathcal{L}_0 \mathcal{L}_1}
	{r_0^3 \left( \frac{\mathcal{L}_0^2}{2} \left(\frac{ a''_0}{r^2_0 a_0} - \frac{a'_0{}^2}{r^2_0 a_0^2}- \frac{2}{r^3_0}\frac{a'_0}{a_0} -  \frac{6}{r^4_0} \right) - m_0^2 \left(\frac{a'_0{}^2}{a_0^2} - \frac{a''_0}{a_0} \right) \right)}\,.
\end{align}

Specifying the background geometry to Schwarzschild geometry by setting $a(r) = (1-\tfrac{r_s}{r})^{-1}$ and considering massless particles, $m=0$, the zeroth order reproduces the universal photon sphere at $r_0 = \frac{3}{2}r_s$, while the influence of a MDR on circular photon orbits in Schwarzschild geometry is
\begin{align}\label{eq:r1photonpert}
	r_1 = \frac{9 r_s^3}{16 \mathcal{L}_0^2}(4h_0 + 3 r_s \partial_r h_0)\,.
\end{align}
In general this correction depends on the angular momentum parameter (or energy) of the particle's trajectory and thus is not universal for all photons anymore, a feature that we will see explicitly for the $\kappa$-Poincar\'e dispersion relations next. In case the perturbation function $h$ is quadratic in the momenta the dependence on the angular momentum vanishes and universality is restored, since then the geometry of spacetime is then determined by a metric, and independent of the momentum coordinates of the cotangent bundle.

The expression \eqref{eq:r1photonpert} gives the quantitative precise correction to the circular photon orbits. Using for example the $\kappa$-Poincar\'e perturbation \eqref{eq:kappaschwpert} gives consistently with previous results obtained in the literature $r_1 = \frac{\mathcal{L}_0}{6}$~\cite{Barcaroli:2017gvg,Pfeifer:2019zhc}, while one can employ \eqref{eq:pertmedium} to investigate the influence of a general medium around a black hole on the photon sphere. For specific media this has been done to all orders, see~\cite[Sec.~IX]{Perlick:2021aok}.

The scale $\epsilon$ at which one expects to observe te deviations depends on the origin of the modification, if it is motivated from quantum gravity phenomenology or scattering effects caused by matter or a medium through which the photons we observe propagate.

\subsubsection{$\kappa$-Poincar\'e}
Applying the algorithm outlined in the beginning of this subsection to the $\kappa$-deformation of Schwarzschild geometry \eqref{eq:kappaschw} gives in the first step from \eqref{eq:dotrsphkappa}
\begin{align}
	e^{\ell\sqrt{a} \mathcal{E}} \frac{p_r}{a} = 0 \Rightarrow p_r = 0\,.
\end{align}
In the second step we can solve $H_\ell(r,\mathcal{E},0,\mathcal{L})=0$, see \eqref{eq:kappaschw}, for $\mathcal{E}$ and find as possible solutions
\begin{align}
	\mathcal{E}_{\pm} = \frac{1}{\ell \sqrt{a}}\ln\left( \frac{r}{r \pm \ell \mathcal{L}}\right)\,.
\end{align}
Finally using that $\dot p_r = \partial_rH_\ell =0$ in \eqref{eq:dotprsphkappa} yields the equation
\begin{align}\label{eq:kappaschwsporb}
	\frac{2 \mathcal{L}}{r \pm \ell \mathcal{L}} \mp \frac{r a'}{\ell a}\ln \left( \frac{r}{r \pm \ell \mathcal{L}}\right) = 0\,.
\end{align}
This transcendental equation determines the circular photon orbits we were looking for, and hence the shadow of circular $\kappa$-deformed Schwarzschild black holes. The main qualitative difference to circular photon orbits in general relativity is that they become angular momentum (or energy) dependent. Hence, the observation or absence of a frequency dependent shadow of a black hole constrains or would be evidence  for a $\kappa$-deformed dispersion relation. The comparison of the photon sphere (or the black hole shadow) in different frequency bands would reveal if the functional dependence on the photon frequency predicted by the $\kappa$-deformed dispersion relation is physically viable, or not.

Using $a = (1-\frac{r_s}{r})^{-1}$ and employing an ansatz $r=r_0 + \ell r_1$ before making a series expansion to first order in $\ell$ around $\ell=0$ in \eqref{eq:kappaschwsporb} yields $r_0 = \frac{3}{2}r_s$ and $r_1 = \pm\frac{\mathcal{L}}{6}$, which are precisely the results already presented in the literature~\cite{Barcaroli:2017gvg,Pfeifer:2019zhc}, which was also already mentioned previously.

\subsubsection{From the photon sphere to the black hole shadow}\label{sec:BHShad}
The precise relation between the radius of the photon sphere, the position of the observer and the angular size of the shadow of a black hole seen by the observer depends on the observer models one employs in the context of the MDR in consideration.

In any case, the crucial relation is that the radius of the photon sphere $r_{\textrm{ph}}$ is the point of closest encounter to the black hole for the photon trajectories, $\gamma(\tau) = (t(\tau),r(\tau),\frac{\pi}{2},\phi(\tau))$, which form the boundary of the shadow. At this point of closest radial distance to the black hole the photon trajectories must satisfy $\dot r=0$, which from $\bar \partial^r H(r_{\textrm{ph}},\mathcal{E},p_r,\mathcal{L}) = 0$ implies for the perturbative Hamiltonian \eqref{eq:mdrpert} that $p_r(r_{\textrm{ph}},\mathcal{E},p_r,\mathcal{L}) = 0+\mathcal{O}(\epsilon)$. Hence, the massless dispersion, $H(r_{\textrm{ph}},\mathcal{E},0,\mathcal{L})=0$, implies a dependence between the constants of motions $\mathcal{E}$, $\mathcal{L}$ and $r_{\textrm{ph}}$
\begin{align}\label{eq:B}
	\frac{\mathcal{L}^2}{\mathcal{E}^2} = a(r_{\textrm{ph}}) r_{\textrm{ph}}^2 - 2 \epsilon \frac{r_{\textrm{ph}}^2}{\mathcal{E}^2}h(r_{\textrm{ph}},\mathcal{E},0,\mathcal{L})\,.
\end{align}
The quantity $B = \frac{\mathcal{L}}{\mathcal{E}}$ is the impact parameter. For dispersion relations which are quadratic in the momenta $B$ itself is independent of $\mathcal{E}$ or $\mathcal{L}$ and depends only on $r_{\textrm{ph}}$. For generic modified dispersion relations the above equation is an implicit relation determining one constant of motion in terms of the other and the photon sphere radius.

Conventionally, when the observer and its measurements are modeled with the zeroth order spacetime metric $g = - a(r)^{-1} \mathrm{d}t^2 + b(r)^{-1} \mathrm{d}r^2 + r^2 (\mathrm{d}\theta^2 + \sin^2\theta \mathrm{d}\phi^2)$, the shadow angle $\alpha$ is given by, see~\cite{Perlick:2021aok}
\begin{align}\label{eq:alphash}
	\cot \alpha = \left(\frac{\sqrt{g_{rr}}}{\sqrt{g_{\phi\phi}}} \frac{\mathrm{d}r}{\mathrm{d}\phi}\right)\bigg|_{r=r_O} = \left(\frac{1}{\sqrt{b(r)}r} \frac{\mathrm{d}r}{\mathrm{d}\phi}\right)\bigg|_{r=r_O}\,,
\end{align}
where $r_O$ is the radial coordinate position of an observer at rest with respect to the black hole. The validity of this expression relies on the assumption that the spacetime geometry the observer experiences is determined by the spacetime metric (in particular the geometry in the $r-\phi$-plane and the notion of angles there), i.e. effects from the modified dispersion relation do not affect the observer.

The term $\tfrac{\mathrm{d}r}{\mathrm{d}\phi} = \tfrac{\dot r}{\dot \phi}$ can be obtained from the Hamilton equations of motion as a function of $r$ and the constants of motion $\mathcal{E}$ and $\mathcal{L}$, where the appearing terms of $p_r$ can be replaced as function of $r,\mathcal{E}$ and $\mathcal{L}$ by solving the massless dispersion relation $H(x,p)=0$ for $p_r(r,\mathcal{E},\mathcal{L})$. Moreover, equation \eqref{eq:B} allows us to replace one of the constants of motion in terms of the other and the photon sphere radius $r_{\textrm{ph}}$. Thus, the angular size of the shadow of the black holes is
\begin{align}\label{eq:shadowSize}
	\cot \alpha(r_O,r_{\textrm{ph}},\mathcal{L}) = \left(\frac{1}{\sqrt{b(r_O)}r_O} \frac{\mathrm{d}r}{\mathrm{d}\phi}(r_O,\mathcal{E}(r_{\textrm{ph}},\mathcal{L}),\mathcal{L})\right)\,.
\end{align}
For quadratic dispersion relations the dependence on $\mathcal{L}$ drops out and the angular width of the shadow is the same for photons of all frequencies. For generic dispersion relations, the shadow size becomes frequency dependent. The precise dependence on $\mathcal{L}$ depends on the modified dispersion relation under consideration, and the dependence of $r_{\mathrm{ph}}$, which itself depends on $\mathcal{L}$, as demonstrated explicitly in equations \eqref{eq:r1} and \eqref{eq:kappaschwsporb}  (instead of $\mathcal{L}$ the whole equation can equally be written as being dependent on $\mathcal{E}$). Following the just outlined algorithm one can predict the functional dependence of the black hole shadow on the photon properties and thus observations of the size of the black hole shadow in different frequencies, and their relative change leads to a falsifiable prediction.

Evaluating \eqref{eq:shadowSize} for general first order modifications of Schwarzschild geometry ($a=b^{-1}=(1-\frac{r_s}{r})^{-1}$), the angular size of the black hole shadow becomes
\begin{align}
	\cot \alpha = \frac{(3 r_{s}-2 r_O) \sqrt{r_O +3 r_{s}}}{3 \sqrt{3} r_{s} \sqrt{r_O-r_{s}}}
	- \epsilon \frac{\sqrt{3} r_O^2 r_{s} \epsilon  (3 (r_{s}-r_O) h(r_O,\mathcal{E}_0,p_{0r},\mathcal{L})  + r_O h(r_{\rm 0ph}, \mathcal{E}_0,0,\mathcal{L})}{\mathcal{L}^2 (2 r_O-3 r_{s}) \sqrt{r_O-r_{s}} \sqrt{r_O+3 r_{s}}}
\end{align}
where we introduced the short hand notations,
\begin{align}
 \mathcal{E}_0^2 = \frac{4 \mathcal{L}^2}{27 r_s^2},\quad p_{0r} = \mathcal{L}\frac{ \sqrt{\frac{4 r_O^3}{r - r_s}- 27 r_s^2}}{3 r_O r_s \sqrt{3-\frac{3 r_s}{r_O}}},\quad r_{\rm 0ph} = \frac{3}{2}r_s\,,
\end{align}
for the zeroth order values of the energy and radial momentum of the photon trajectory that forms the boundary of the shadow. Note that the observer is always outside the photon sphere, i.e. $r_O>3/2 r_s$ and thus all appearing roots are real. The zeroth order nicely gives the usual formula for the size of the Schwarzschild black hole shadow determined on the basis of general relativity, see~\cite{Perlick:2021aok}, which often is expressed in terms of $\sin \alpha^2 = (\cot \alpha ^2 + 1)^{-1}$.

For the $\kappa$-Poincar\'e dispersion relation we find
\begin{align}
		\cot \alpha = \frac{1}{\ell \mathcal{L}}
	r_{\rm ph}^{-\frac{\sqrt{a(r_O)}}{\sqrt{a(r_{\rm ph})}}} 
	\sqrt{r_O^2 \left(r_{\rm ph}^{\frac{\sqrt{a(r_O)}}{\sqrt{a(r_{\rm ph})}}}-(\ell \mathcal{L}+r_{\rm ph})^{\frac{\sqrt{a(r_O)}}{\sqrt{a(r_{\rm ph})}}}\right)
	^2-\ell^2 \mathcal{L}^2 r_{\rm ph}^{\frac{2 \sqrt{a(r_O)}}{\sqrt{a(r_{\rm ph})}}}}\,,
\end{align}
where the circular photon orbit $r_{\rm ph}=r_{\rm ph}(\mathcal{L})$ is determined by \eqref{eq:kappaschwsporb}.

One word of caution is an order here. In the whole argument above we assumed a general relativistic observer model. This is not valid in the context of DSR, Finslerian or Hamiltonian observer models. However in all of these theories the observer models on curved spacetimes are not completely constructed yet. In particular an observers measurement of angles has, to our knowledge, not yet been formulated in these tangent/cotangent bundle geometries. For a self-consistent theoretical descriptions of the propagation effects and observer measurements in the context of quantum gravity phenomenology this is a crucial point in our opinion. We expect that in such models the influence of the MDR is not only visible in the $\frac{\mathrm{d}r}{\mathrm{d}\phi}$ term in equation \eqref{eq:alphash}, but will also appear in the geometric term $\frac{\sqrt{g_{rr}}}{\sqrt{g_{\phi\phi}}}$, which comes from the pull back of the spacetime geometry to the $r-\phi$ plane. How this works precisely in the geometric frameworks beyond metric spacetime geometry is an open question and currently under investigation.

\subsection{Shapiro delay}\label{ssec:shapiro}
The Shapiro delay is the difference between the time of flight of a light-signal in the presence and the absence of a gravitating central object. The comparison of the derived Shapiro delay with the experimental measurements leads to test of the gravitational field in the weak field approximation~\cite{Possel:2021vxj}, for example with lunar laser ranging~\cite{Merkowitz:2010kka} as well as the strong gravity regime with pulsars orbiting black holes~\cite{Hackmann:2018fmk}.

As for the circular photon orbits we will find that the Shapiro delay will become dependent on the energy of the photons and no longer be universal, which again is the eperimental signature to constrain or find evidence for MDRs.

To derive this difference we consider the following situation in the equatorial plane. A light ray is emitted at a position $r = r_e$, propagates towards the central gravitating object, reaches a closest encounter at $r=r_c$, gets reflected at a mirror at $r=r_m$ and travels back to $r_e$. Then the total time of flight is given by
\begin{align}
	\Delta T = 2 \left( \int_{r_c}^{r_e} \mathrm{d}\bar r \frac{\mathrm{d} t}{\mathrm{d} r} + \int_{r_c}^{r_m} \mathrm{d} \bar r \frac{\mathrm{d} t}{\mathrm{d} r} \right)\,.
\end{align}
The function $\mathrm{d} t/\mathrm{d} r$ can directly be obtained from the Hamilton equations of motion.

First note that at the point of closest encounter $r=r_c$ of the light signal to the central gravitating object $\dot r = 0$ and thus \eqref{eq:dotrsph} implies that $\bar\partial^rH=0$. This equation can be used to determine $p_r = p_r(r_c, \mathcal{E},\mathcal{L})$. Next we can solve the dispersion relation $H(r_c, \mathcal{E}, p_r(r_c, \mathcal{E},\mathcal{L}),\mathcal{L})=0$ at $r_c$ to obtain $\mathcal{L} = \mathcal{L}(r_c,\mathcal{E},p_r(r_c))$. Along the whole light trajectory we can then solve the dispersion relation $H(r, \mathcal{E}, p_r,\mathcal{L}(r_c,\mathcal{E},p_r(r_c))=0$ for $p_r = p_r(r,\mathcal{E},r_c)$. Finally we can derive the essential ingredient for the time delay using equations \eqref{eq:dottsph} and \eqref{eq:dotrsph} in as function of the energy, the radius of closest encounter and the radial position of the particle as
\begin{align}\label{eq:dtdr}
	\frac{\mathrm{d} t}{\mathrm{d} r} = \frac{\dot t}{\dot r} = \frac{\bar\partial^t H(r, \mathcal{E},p_r(r,\mathcal{E},r_c),\mathcal{L}(r_c,\mathcal{E},p_r(r_c) )}{\bar\partial^r H(r, \mathcal{E},p_r(r,\mathcal{E},r_c),\mathcal{L}(r_c,\mathcal{E},p_r(r_c) )} = \frac{\mathrm{d} t}{\mathrm{d} r}(r,r_c,\mathcal{E})\,.
\end{align}

\subsubsection{General first order modified dispersion relations}\label{ssec:shapMDR}
To study the Shapiro delay for general first order MDRs we follow the algorithm outlined in the introduction to this subsection. First,  at the point of closest encounter $r_c$ to the gravitating central mass, where $\dot r =0$, equation \eqref{eq:dotrpert} implies that $p_r$ has only a first order contribution
\begin{align}\label{eq:prcritpert}
	p_r(r_c, \mathcal{E},\mathcal{L}) = p_{r_c} = - \epsilon \frac{\bar\partial^r h(r_c,\mathcal{E},0,\mathcal{L})}{b_c}\,,
\end{align}
where we set $b_c = b(r_c)$ and use $a_c = a(r_c)$ later. From the dispersion relation \eqref{eq:mdrpert} one then easily concludes, setting $h(r_c,\mathcal{E},0,\mathcal{L}) = h_c$,
\begin{align}\label{eq:Lcritpert}
	\mathcal{L}(r_c,\mathcal{E})^2 = \mathcal{E}^2 a_c r_c^2 - \epsilon 2 r_c^2 h_c\,.
\end{align}
Using \eqref{eq:prcritpert} and \eqref{eq:Lcritpert} in the dispersion relation \eqref{eq:mdrpert} yields, to first order in $\epsilon$,
\begin{align}\label{eq:prpert}
	p_r(r,\mathcal{E},r_c)
	&= p_{0r} + \epsilon \frac{1}{b\ p_{0r}} \left( \frac{r_c^2}{r^2}h(r_c,\mathcal{E},0,\mathcal{L}(r_c,\mathcal{E})) - h(r,\mathcal{E},p_{0r},\mathcal{L}(r_c,\mathcal{E}))\right)\,,
\end{align}
where $p_{0r}= \pm \mathcal{E}\sqrt{\frac{a}{b}}\sqrt{1 - \frac{ a_c r_c^2}{a r^2}}$. Now we can combine all results to evaluate \eqref{eq:dtdr}:
\begin{align}
	\frac{\mathrm{d} t}{\mathrm{d} r}( r,r_c,\mathcal{E})
	&= \frac{- a \mathcal{E} + \epsilon \bar\partial^t h(r,\mathcal{E}, p_r ,\mathcal{L})}{b p_r + \ell \bar\partial^r h(r,\mathcal{E}, p_r, \frac{\pi}{2},\mathcal{L})} \\
	&= - \frac{a\mathcal{E}}{b p_{0r}} + \epsilon \left( - \frac{a\mathcal{E} (h(r,\mathcal{E},p_{0r},\mathcal{L}) - \frac{r_c^2}{r^2}h(r_c,\mathcal{E},0,\mathcal{L}))}{b^2 p_{0r}^3}  + \frac{\bar\partial^t h(r,\mathcal{E}, p_{0r} ,\mathcal{L})}{b p_{0r}} + \frac{a \mathcal{E} \bar\partial^r h(r,\mathcal{E}, p_{0r},\mathcal{L})}{b^2 p_{0r}^2} \right)
\end{align}
Specifying to Schwarzschild geometry $b=a^{-1}$ yields
\begin{align}\label{eq:dtdr1storder}
	\frac{\mathrm{d} t}{\mathrm{d} r}(r,r_c,\mathcal{E}) = -\frac{a^2 \mathcal{E}}{p_{0r}}
	+ \epsilon \left( \frac{a \bar\partial^t h(r,\mathcal{E},p_{0r},\mathcal{L})}{p_{0r}} +   \frac{ \mathcal{E} a^3 \bar\partial^r h(r,\mathcal{E},p_{0r},\mathcal{L})}{p_{0r}^2} - \frac{\mathcal{E} a^3 (h(r,\mathcal{E},p_{0r},\mathcal{L}) - \frac{r_c^2}{r^2} h(r_c,\mathcal{E},0,\mathcal{L}))}{p_{0r}^3}\right)
\end{align}
In general it depends on the perturbation function of interest if the integrals can be evaluated or not.

The time of flight of a photon and the Shapiro delay can be derived from the above equation in the same way as we did it for the all order $\kappa$-Poincar\'e dispersion relation in the previous section. For any given perturbation function $h$ one derives
\begin{align}
	\Delta T(r_c,R)
	&= \int_{r_c}^R \mathrm{d}\bar r \frac{\mathrm{d} t}{\mathrm{d} r}(\bar r,r_c,\mathcal{E})
\end{align}
which yields the Shapiro delay
\begin{align}
	\frac{1}{2} \Delta T_{\textrm{pert-Shapiro}} &= (\Delta T(r_c,R_e) + \Delta T(r_c,R_m)) -  \left( \sqrt{R_e^2-r_c^2} + \sqrt{R_m^2-r_c^2} \right)\,.
\end{align}
The first order $\kappa$-Poincar\'e result can be obtained by employing the perturbation function \eqref{eq:kappaschwpert} with $a = (1-r_s/r)^{-1}$ and performing linearisation in $r_s$ to get
\begin{align}
	\Delta T(r_c,R) = \sqrt{R^2-r_c^2} (1 - \ell \mathcal{E}) + r_s \left( 1 - \frac{3}{2} \ell \mathcal{E}\right)
	\left(\frac{1}{2} \frac{\sqrt{R - r_c}}{\sqrt{R + r_c}} +  \ln\left(\frac{R+\sqrt{R^2-r_c^2}}{r_c}\right)\right)\,.\label{eq:trcRL}
\end{align}
which is identical to what one obtains from linearising the all order $\kappa$-Poincar\'e expression, which we derive in \eqref{eq:TrcRkappa}, directly in $\ell$.

\subsubsection{$\kappa$-Poincar\'e}\label{ssec:shapkappa}
For the $\kappa$-Poincar\'e deformation of Schwarzschild geometry \eqref{eq:kappaschw} the algorithm outline in the beginning of this subsection yields the following results.

First, at the point of closest encounter we have $\dot r = 0$, which implies from \eqref{eq:dotrsphkappa} that $p_r(r_c,\mathcal{E},\mathcal{L}) = 0$. Hence $H_\ell(r_c, \mathcal{E}, p_r(r_c, \mathcal{E},\mathcal{L}),\mathcal{L})= H_\ell(r_c, \mathcal{E}, 0,\mathcal{L})= 0$ can be solved for $\mathcal{L}$ as
\begin{align}\label{eq:Lofrc}
	\mathcal{L}(r_c,\mathcal{E})^2 = \frac{4}{\ell^2}r_c^2 e^{- \ell \sqrt{a_c}\mathcal{E}}\sinh\left(\frac{\ell}{2}\sqrt{a_c}\mathcal{E}\right)^2\,,
\end{align}
where $a_c = a(r_c)$. Second, we solve $H_\ell(r, \mathcal{E}, p_r,\mathcal{L}(r_c,\mathcal{E}))=0$ for $p_r = p_r(r,\mathcal{E},r_c)$ and find
\begin{align}\label{eq:profrc}
	p_r(r,\mathcal{E},r_c)^2 = \frac{4}{\ell^2} a \left(  e^{-\ell \sqrt{a} \mathcal{E}} \sinh\left(\frac{\ell}{2}\sqrt{a}\mathcal{E}\right)^2 - \frac{r_c^2}{r^2}e^{-\ell \sqrt{a_c} \mathcal{E}} \sinh\left(\frac{\ell}{2}\sqrt{a_c}\mathcal{E}\right)^2\right)\,.
\end{align}
Third we combine all these equations to evaluate \eqref{eq:dtdr} with help of the equations \eqref{eq:dottsphkappa} and \eqref{eq:dotrsphkappa} to obtain the desired expression
\begin{align}
	\frac{\mathrm{d} t}{\mathrm{d} r}(r,r_c,\mathcal{E}) = \frac{r a e^{-2\ell \sqrt{a}\mathcal{E}}(e^{\ell \sqrt{a}\mathcal{E}} - 1)}{\sqrt{e^{-2\ell \sqrt{a}\mathcal{E}}(e^{\ell \sqrt{a}\mathcal{E}} - 1)^2 r^2 - e^{-2\ell \sqrt{a_c}\mathcal{E}}(e^{\ell \sqrt{a_c}\mathcal{E}} - 1)^2 r_c^2}}
\end{align}
Integrating this function for general $a=a(r)$ without further assumptions is not possible.

To derive an explicit expression for the Shapiro delay we consider Schwarzschild geometry $a = (1-\frac{r_s}{r})^{-1}$ and linearize in the Schwarzschild radius $r_s$, i.e.\ we assume that the light ray we consider satisfies $r \gg r_s$,
\begin{align}
	\frac{\mathrm{d} t}{\mathrm{d} r}(r,r_c,\mathcal{E}) =  \frac{e^{-\ell \mathcal{E}}}{\sqrt{r^2-r_c^2}}\left( r + r_s \frac{ ( 1 - e^{\ell \mathcal{E}} ) (\ell \mathcal{E} - 2 ) r + ( 2 ( \ell \mathcal{E} - 1) + e^{\ell \mathcal{E}} (2 - \ell \mathcal{E}) ) r_c}{2 (e^{\ell\mathcal{E}} - 1 )(r + r_c)} \right)\,.
\end{align}

This function can easily be integrated and we find for  the travel time of light between an arbitrary point $R>r_c$ and $r_c$
\begin{align}\label{eq:TrcRkappa}
	\Delta T(r_c,R)
	&= \int_{r_c}^R \mathrm{d}\bar r \frac{\mathrm{d} t}{\mathrm{d} r}(\bar r,r_c,\mathcal{E}) \nonumber \\
	&= e^{- \ell \mathcal{E}}\left( \sqrt{R^2 - r_c^2} + r_s \frac{\ell \mathcal{E}}{2 (e^{\ell \mathcal{E}} - 1)} \frac{\sqrt{R - r_c}}{ \sqrt{R + r_c}}
	+  r_s  \frac{( 2- \ell \mathcal{E} )}{2} \ln\left(\frac{R+\sqrt{R^2-r_c^2}}{r_c}\right) \right)\,.
\end{align}
The Shapiro delay then is given by
\begin{align}
	\frac{1}{2}\Delta T_{\kappa-Shapiro} =  \Delta T(r_c,r_e) + \Delta T(r_c,r_m) - ( \Delta T(r_c,r_e) + \Delta T(r_c,r_m)  )|_{r_s=0}\,.
\end{align}
In each part of the light propagation, between $r_e$ and $r_c$ and between $r_c$ and $r_m$ the important terms are
\begin{align}\label{eq:TrcRkappaTerms}
	\Delta T(r_c,r_e) - (\Delta T(r_c,r_e)|_{r_s=0})
	= r_s e^{- \ell \mathcal{E}}\left(  \frac{\ell \mathcal{E}}{2 (e^{\ell \mathcal{E}} - 1)} \frac{\sqrt{r_e - r_c}}{\sqrt{ r_e + r_c}} + \frac{( 2- \ell \mathcal{E} )}{2} \ln\left(\frac{r_e +\sqrt{r_e^2-r_c^2}}{r_c}\right) \right)\,.
\end{align}
As announced in the beginning of this section, the Shapiro delay becomes dependent on the energy parameter of the light trajectory.

The energy dependent time of arrival of photons is investigated in the context of quantum gravity phenomenology in a cosmological setting for a long time~\cite{Pfeifer:2018pty,Terzic:2021rlx,Bolmont:2022yad,Acciari:2020kpi}. The results presented here are the beginning of extending the search for such effects from MDRs around black holes. The Shapiro delay has been measured to post Newtonian order with help of the Cassini spacecraft~\cite{Bertotti:2003rm,Will:2014kxa} and is necessary to describe pulsar timing in the vicinity of a black hole correctly~\cite{Hackmann:2018fmk}. Thus, if one would prepare a kind of Cassini tracking experiment with different photon frequencies or find two pulsars, quasars (or even sources which emit gamma rays), in an orbit around a black hole (for example the black hole at the center of our galaxy) in comparable distance which emit photons at different energies, a different time of arrival of the photons detected here can be compared against the time delay formula derived above. However, the authors are currently not aware about any experiments dedicated to the search of a frequency dependence in these observables.



\subsection{Light deflection}\label{ssec:light}
We derive the deflection angle $\Delta \phi$ which characterizes the deviation from a lightlike solution of the Hamilton equations of motion in the presence of a gravitating central mass from a straight line, which would be the lightlike solution of the Hamilton equations of motion in the absence of a gravitating mass.  This derivation complements  the existing results on weak gravitational lensing from MDRs in the literature~\cite{Glicenstein:2019rzj} and is a step towards a general gravitational lensing equation from  MDRs, which is work in progress.

We consider a light pulse in the equatorial plane that propagates from infinitely far away towards the source of the gravitational field, reaches a point of closest distance $r=r_c$, and then propagates away to infinite distance. The deflection angle in this process is the deviation of the light trajectories $\phi$ coordinate from the value $\pi$ after passing to infinite distance.

To derive the deflection angle we follow a similar algorithm as for the time delay. The quantities $p_r = p_r(r_c, \mathcal{E},\mathcal{L})$ and $\mathcal{L} = \mathcal{L}(r_c,\mathcal{E},p_r(r_c))$ as well as $p_r = p_r(r,\mathcal{E},r_c)$ are determined as outlined above \eqref{eq:dtdr}. Then using equations \eqref{eq:dotphisph} and \eqref{eq:dotrsph} in
\begin{align}\label{eq:dphidr}
	\frac{\mathrm{d}\phi}{\mathrm{d} r} = \frac{\dot \phi}{\dot r} = \frac{\bar\partial^\phi H(r, \mathcal{E},p_r(r,\mathcal{E},r_c),\mathcal{L}(r_c,\mathcal{E},p_r(r_c) )}{\bar\partial^r H(r, \mathcal{E},p_r(r,\mathcal{E},r_c),\mathcal{L}(r_c,\mathcal{E},p_r(r_c) )} = \frac{\mathrm{d}\phi}{\mathrm{d} r}(r,r_c,\mathcal{E})\,,
\end{align}
we can determine the light deflection angle $\Delta\phi$ by integrating this quantity
\begin{align}
	\pi + \Delta\phi = 2 \int_{r_c}^\infty \mathrm{d}\bar r \frac{\mathrm{d}\phi}{\mathrm{d} r}(\bar r)\,.
\end{align}

\subsubsection{General first order modified dispersion relations}\label{ssec:lightMDR}
We proceed analogous as in Section \ref{ssec:shapMDR}, where we derived the Shapiro delay for perturbations of a general first order modification of the general relativistic dispersion relation. The equations \eqref{eq:prcritpert}, \eqref{eq:Lcritpert} and \eqref{eq:prpert} yield the quantities $\mathcal{L}(r_c,\mathcal{E},p_r(r_c))$ and $p_r(r,\mathcal{E},r_c)$, which we use in \eqref{eq:dphidr} to obtain
\begin{align}\label{eq:dphidrL}
	\frac{\mathrm{d}\phi}{\mathrm{d} r}( r,r_c,\mathcal{E})
	&= \frac{\dot \phi}{\dot r}
	= \frac{ \frac{\mathcal{L}}{r^2} + \epsilon \partial_w h(r,\mathcal{E}, p_r,\mathcal{L})}{b p_r + \epsilon \bar\partial^r h(r,\mathcal{E}, p_r,\mathcal{L})}
	= \frac{\mathcal{E} r_c \sqrt{a_c} }{b p_{0r} r^2} \nonumber \\
	&+ \epsilon \Bigg(\frac{\partial_w h(r,\mathcal{E}, p_r,\mathcal{L})  - \frac{r_c}{r} \frac{h(r_c,\mathcal{E},0,\mathcal{L})}{r \mathcal{E} \sqrt{a_c}}}{b p_{0r}}
	- \frac{r_c \mathcal{E} \sqrt{a_c} \bar\partial^r h(r,\mathcal{E}, p_r,\mathcal{L})}{b^2 p_{0r}^2 r^2} \nonumber\\
	&+ \frac{r_c \mathcal{E} \sqrt{a_c} (h(r,\mathcal{E}, p_r,\mathcal{L}) - \frac{r_c^2}{r^2}h(r_c,\mathcal{E},0,\mathcal{L}))}{r^2 b^2 p_{0r}^3}\Bigg)\,.
\end{align}
Specifying to Schwarzschild geometry $b=a^{-1}$ yields no significant simplification of the above equation, nor a more compact form, which is why we do not display the equation explicitly.

The deflection angle of a light ray by the central mass is given by integrating this expression as
\begin{align}\label{eq:lightdef}
	\pi + \Delta\phi = 2 \int_{r_c}^\infty \mathrm{d}\bar r \frac{\mathrm{d}\phi}{\mathrm{d} r}(\bar r,r_c,\mathcal{E})\,.
\end{align}

One obtains the first order $\kappa$-Poincar\'e result \eqref{eq:DphikappaL} from \eqref{eq:lightdef} by employing the perturbation function \eqref{eq:kappaschwpert} and performing linearisation in $r_s$ in \eqref{eq:dphidrL}.

\subsubsection{$\kappa$-Poincar\'e}\label{ssec:lightkappa}
For the Schwarzschild $\kappa$-Poincar\'e Hamiltonian \eqref{eq:kappaschw}, the quantities $\mathcal{L}(r_c,\mathcal{E},p_r(r_c))$ and $p_r(r,\mathcal{E},r_c)$ are given by equation \eqref{eq:Lofrc} and \eqref{eq:profrc}. Using these \eqref{eq:dotphisphkappa} and \eqref{eq:dotrsphkappa} in \eqref{eq:dphidr} gives
\begin{align}
	\frac{\mathrm{d} \phi}{\mathrm{d} r}(r,r_c,\mathcal{E}) = \frac{r_c}{r }\frac{ e^{-\ell \sqrt{a_c}\mathcal{E}} \sqrt{a} (e^{\ell \sqrt{a_c}\mathcal{E}} - 1)}{\sqrt{e^{-2\ell \sqrt{a}\mathcal{E}}(e^{\ell \sqrt{a}\mathcal{E}} - 1)^2 r^2 - e^{-2\ell \sqrt{a_c}\mathcal{E}}(e^{\ell \sqrt{a_c}\mathcal{E}} - 1)^2 r_c^2}}\,.
\end{align}
To perform the integration for the light deflection we insert $a = (1-\frac{r_s}{r})^{-1}$ and linearize in $r_s$
\begin{align}
	\frac{\mathrm{d} \phi}{\mathrm{d} r}(r,r_c,\mathcal{E}) = \frac{1}{r \sqrt{ r^2- r_c^2 }}\left( r_c + r_s \frac{r_c (r + r_c)( e^{\ell \mathcal{E}} - 1 ) + \ell \mathcal{E} r^2}{2 r ( e^{\ell \mathcal{E}} - 1 ) (r + r_c)}\right) \,.
\end{align}
This function can easily be integrated and we find the primitive
\begin{align}
	\int \mathrm{d}\bar r \frac{\mathrm{d}\phi}{\mathrm{d} r}(\bar r,r_c,\mathcal{E})= \tan^{-1}\left( \frac{\sqrt{r^2 - r_c^2}}{r_c} \right) + r_s \frac{\sqrt{r-r_c}}{\sqrt{r+r_c}} \frac{( r  ( e^{\ell \mathcal{E}} - 1 + \ell \mathcal{E} ) + r_c  ( e^{\ell \mathcal{E}} - 1) )}{ 2 r r_c  ( e^{\ell \mathcal{E}} - 1)}\,,
\end{align}
from which we obtain the light deflection angle
\begin{align}\label{eq:Dphikappa}
	\Delta\phi = \left(2 \int_{r_c}^\infty \mathrm{d}\bar r \frac{\mathrm{d}\phi}{\mathrm{d} r}(\bar r,r_c,\mathcal{E})  \right) - \pi = 2 \left( \frac{r_s}{r_c} \frac{(e^{\ell \mathcal{E}} - 1 + \ell \mathcal{E})}{2 (e^{\ell \mathcal{E}} - 1)} \right)\,.
\end{align}
To first order in the $\kappa$-Poincar\'e deformation parameter $\ell$ it simplifies  to
\begin{align}\label{eq:DphikappaL}
	\Delta\phi = \frac{2 r_s}{r_c} - \ell \mathcal{E} \frac{r_s}{2 r_c}\,.
\end{align}

\section{Conclusion and Outlook}\label{sec:conc}
Due to new instruments and new observation channels such as multi-messenger, the search for traces of quantum gravity becomes a more and more realistic task~\cite{Addazi:2021xuf}. Their possible manifestation in terms of a MDR needs a precise mathematical formulation on curved spacetimes. With this article we presented how this can be achieved for spherically symmetric physical systems.

We derived the explicit expressions for the photon sphere, the Shapiro delay and the deflection angle of light for photons that are subject to either a general first order modification of the general relativistic dispersion relation or the the $\kappa$-Poincar\'e dispersion relation in the bicrossproduct basis on curved spacetimes. This investigation extends the efforts to find signals or constrain MDRs with astrophysical observations to spherically symmetric systems, which approximate the more realistic axial symmetric situation of rotating astrophysical objects.

The main difference between the general relativistic result and the one obtained for the MDRs, is that these characteristic quantities become generically energy or angular momentum dependent. We quantified this dependence for the perturbative MDRs in the equations \eqref{eq:r1}, \eqref{eq:dtdr1storder} and \eqref{eq:dphidrL}, as well as for the $\kappa$-Poincar\'e dispersion relations in equations \eqref{eq:kappaschwsporb}, \eqref{eq:TrcRkappa} and \eqref{eq:Dphikappa}. For the perturbative MDRs the equations \eqref{eq:r1}, \eqref{eq:dtdr1storder} and \eqref{eq:dphidrL} can  be used to derive the desired quantities for any perturbation function one may consider and one can identify precise conditions on the MDR such that they stay frequency independent; a feature that is certainly the case for modification that are quadratic in the 4-momentum of the particles, however these may not be the only possibilities.

Constraints or evidences for MDRs can be found from astrophysical observables in spherical symmetry which are influenced by the photon sphere, the Shapiro delay and the deflection angle of light such as: black hole shadows, as we outline in Section \ref{sec:BHShad}, Cassini tracking and Pulsar timing with different frequencies, as we commented in Section \ref{ssec:shapkappa} and in gravitational lensing images.

In case of a detection of such an effect, its origin needs to be discussed with help of the scale and magnitude at which it appears to disentangle possible effects from a medium in the path of the particle between its origin and the detector and possible quantum gravity effects.

For quantum gravity effects it is expected that the perturbation parameter $\ell$ is of the order of or identical to the tiny Planck length $\ell \sim 10^{-35}\mathrm{m}$. In order to have a chance to detect effects which are suppressed by such a scale amplifications are needed. In the study of time delays on cosmological scales these are the cosmological distances over which the effect accumulates to a hopefully measurable size. In spherical symmetry, in general such an amplifier does not exist.

However, for the Shapiro delay we see from equation \eqref{eq:trcRL} that possible amplification factors can be the the Schwarzschild radius, since it multiplies the energy dependent modifications. The most massive black holes which have been detected have a mass of $10^{10}$ solar masses, see~\cite{Shemmer:2004ph,Mehrgan_2019}, and thus a Schwarzschild radius is of the order $10^{13}\mathrm{m}$, which surely is way smaller compared to cosmological distances. When multiplied with the Planck length the energy dependence of the Shapiro delay goes with $10^{-25} \mathrm{s}/\mathrm{eV}$ multiplied with the energy parameter of the photon trajectories. Thus, in order to set a realistic bound on the quantum gravity scale $\ell$ from the absence of a frequency dependence, either a measurement precision for time delays of this order is required, or very high energetic photons are needed. This might change when one considers non-asymptotically flat spherical spacetime geometries as background such as Schwarzschild-(anti)De-Sitter spacetimes, which combine cosmological and black hole effects.

An ongoing and future research project is to use the results presented here to invent a specific experimental setup dedicated to the detection of the predicted frequency dependencies in the order of magnitude needed.

The canonical next step in the systematic analysis of the influence of MDRs is the derivation of a general lensing equation and the derivation of observables in axial symmetry to investigate the influence of the rotation paramater of the black hole on the energy dependent corrections emerging from modified dispersion relations.

Another interesting possibility to be investigated in the future is that the curvature of spacetime sources the MDR, i.e.\ to consider perturbation functions of the type $h = R^{ab}(x)p_c p_d$, $h = R(x) f(x,p)$ or $h = R^{abcd}(x)R_{abcd}(x) f(x,p)$. Such terms emerge for example from renormalization of quantum field theories on curved spacetimes, as has been pointed out long ago in the case of QED~\cite{Drummond:1979pp}. Moreover such curvature induced MDRs have not been investigated in the literature yet and realize the idea that MDRs curved are caused by (quantum)-gravity.

\section*{Acknowledgments}
C.P. was funded by the Deutsche Forschungsgemeinschaft (DFG, German Research Foundation) - Project Number 420243324 and acknowledges the excellence cluster EXC-2123 QuantumFrontiers - 390837967 funded by the Deutsche Forschungsgemeinschaft (DFG, German Research Foundation) under Germany's Excellence Strategy. D.L. and M.H. acknowledge financial support by the Estonian Ministry for Education and Science through the Personal Research Funding Grants PSG489 and PRG356 and the European Regional Development Fund through the Center of Excellence TK133 ``The Dark Side of the Universe''. The authors would like to acknowledge networking support by the COST Action CA18108 (QGMM).

\end{document}